\documentclass[aps,prb,twocolumn,floatfix]{revtex4}
\usepackage{graphicx}

\begin{document}

\title{How tight is the Lieb-Oxford bound?}
\author{Mariana M. Odashima and K. Capelle}
\email{capelle@ifsc.usp.br}
\affiliation{Departamento de F\'{\i}sica e Inform\'atica\\
Instituto de F\'{\i}sica de S\~ao Carlos\\
Universidade de S\~ao Paulo\\
Caixa Postal 369, S\~ao Carlos, 13560-970 SP, Brazil}

\date{\today}

\begin{abstract}
Density-functional theory requires ever better exchange-correlation (xc)
functionals for the ever more precise description of many-body effects on 
electronic structure. Universal constraints on the xc energy are important 
ingredients in the construction of improved functionals. Here we investigate 
one such universal property of xc functionals: the Lieb-Oxford lower bound on 
the exchange-correlation energy, $E_{xc}[n] \ge -C \int d^3r\,n^{4/3}$, where 
$C\leq C_{LO}=1.68$. To this end, we perform a survey of available exact or
near-exact data on xc energies of atoms, ions, molecules, solids, and some
model Hamiltonians (the electron liquid, Hooke's atom and the Hubbard model). 
All physically realistic density distributions investigated are consistent 
with the tighter limit $C \leq 1$. For large classes of systems one can obtain
class-specific (but not fully universal) similar bounds. The Lieb-Oxford bound 
with $C_{LO}=1.68$ is a key ingredient in the construction of modern xc 
functionals, and a substantial change in the prefactor $C$ will have 
consequences for the performance of these functionals.
\end{abstract}


\maketitle

\newcommand{\be}{\begin{equation}}
\newcommand{\ee}{\end{equation}}
\newcommand{\bi}{\bibitem}
\newcommand{\la}{\langle}
\newcommand{\ra}{\rangle}
\newcommand{\ua}{\uparrow}
\newcommand{\da}{\downarrow}
\newcommand{\bea}{\begin{eqnarray}}
\newcommand{\eea}{\end{eqnarray}}
\renewcommand{\r}{({\bf r})}
\newcommand{\rp}{({\bf r'})}

\section{\label{intro}Introduction}
Any numerical calculation of the electronic structure of matter that uses 
density-functional theory (DFT) employs an approximate exchange-correlation
(xc) functional.\cite{kohnrmp,dftbook,parryang} Further progress in DFT thus 
depends crucially on the development of ever better density functionals. A 
most important ingredient in this quest for better functionals is the small, 
but increasing, list of exact properties of and constraints on the universal 
xc functional.\cite{dftbook,parryang,perdewreview} 

In particular, this functional, $E_{xc}$, is known
to satisfy the following inequalities\cite{perdewreview,lp93}
\be
0 \ge E_{xc}[n] \ge B[n] \ge -C \int d^3r\,n^{4/3},
\label{ineq}
\ee
where $C$ is a universal constant,
$B[n]=\lim_{\gamma\to 0} \gamma^{-1}E_{xc}[n_\gamma]$, and 
$n_\gamma$ is the scaled density $n_\gamma\r=\gamma^3n(\gamma{\bf r})$. While 
the first inequality, providing an upper bound on $E_{xc}[n]$, is an immediate 
consequence of the variational principle, the second and third, providing 
lower bounds, are more complex. The second inequality, the Levy-Perdew 
bound,\cite{lp93} is based on scaling arguments. It contains $E_{xc}$ on 
the left and on the right, and thus provides a consistency test for any given 
approximation to $E_{xc}[n]$. 

The third inequality is a remarkable result due to Lieb and 
Oxford,\cite{lopaper}, who established the form of the bound and obtained 
the value $C_{LO}=1.68$ as an upper limit\cite{footnote1} of the prefactor 
$C$. The present work is mostly concerned with this Lieb-Oxford
bound, although a numerical comparison with the Levy-Perdew bound will
also be given. In terms of the local-density approximation (LDA) to the
exchange energy,
\be
E_x^{LDA}[n]= -\frac{3}{4}\left(\frac{3}{\pi}\right)^{1/3} \int d^3r\,n^{4/3},
\ee
the LO bound can also be written as\cite{perdewreview,lp93}
\be
E_{xc}[n] \ge \lambda_{LO} E^{LDA}_x[n],
\label{loalternat}
\ee
where $\lambda_{LO}=1.354 C_{LO}=2.275$. The analysis below is couched in 
terms of $\lambda$.

The Lieb-Oxford lower bound on the $xc$ energy is one of not many
exactly known properties of the universal $xc$ functional. Similarly to other
such properties, is has been used as a constraint in the construction of
approximations to this functional.\cite{perdewreview} It is satisfied, e.g., 
by the LDA,\cite{pw92} the PBE generalized-gradient approximation\cite{pbe} 
(GGA) and the TPSS meta-GGA.\cite{tpss} On the other hand, earlier 
GGAs\cite{pw86} and semiempirical functionals containing fitting 
parameters\cite{b88,lyp} are not guaranteed to satisfy the bound for 
all possible densities.

Note that Eq.~(\ref{loalternat}) is a bound in the mathematical sense, i.e., 
$E_{xc}[n]$ can never be more negative than $\lambda_{LO} E^{LDA}_x[n]$.  
It is, however, not clear from the inequality itself if $\lambda_{LO}$ is 
the smallest possible value of the prefactor, i.e., if the bound can be 
tightened or not. Indeed, Chan and Handy\cite{handylo} have revisited the
original calculation of Lieb and Oxford, and obtained the
slightly tighter bound $\lambda_{CH}=2.2149$ (or $C_{CH}=1.6358$).

Independently of the question whether the bound can be tightened 
mathematically, it is not clear if nature actually makes use of the entire 
range of values of $E_{xc}$ allowed by the bound, and neither how distant
specific classes of actual physical systems are from the mathematical maximum.
To put these issues in clearer focus, note that for any actual density
$n\r$ one can, in principle, evaluate the density functionals $E_{xc}$
and $E_x^{LDA}$ on this density, and calculate the ratio
\be
\frac{E_{xc}[n]}{E_x^{LDA}[n]}=:\lambda[n],
\label{lambdan}
\ee
which measures the weight of LDA exchange relative to the full 
exchange-correlation energy.
The resulting value of $\lambda[n]$ must be smaller than or equal to
$\lambda_{LO}=2.275$, for any $n$, but it is not {\em a priori} clear by how 
much, and neither what the variations of $\lambda$ over different classes of 
systems are. 

Such information is not easy to obtain, since the
definition (\ref{lambdan}) requires knowledge of the exact $xc$ energy in
the numerator and of the exact density $n\r$ in the numerator and the
denominator. This knowledge is, in general, not available.
There are, however, certain classes of systems for which near-exact $xc$
energies and densities are available, e.g., from quantum Monte Carlo (QMC) or
configuration interaction (CI) calculations. In this work we present a
survey of avaliable such data for large and distinct classes of systems,
and confront the results with the Lieb-Oxford bound.

Section~\ref{atoms} deals with real atoms, Sec.~\ref{hooke} with
Hooke's atom, Sec.~\ref{ions} with ions, Sec.~\ref{molecules} with
a few molecules and Sec.~\ref{liquid} with the homogeneous electron liquid.
Sec. \ref{sum} synthesizes the empirical analysis of Secs. \ref{atoms} to
\ref{liquid} in the form of two conjectures. Readers who do not want to 
go through the details of the analysis of different types of systems can go
right to the conclusions, where all essential results are summarized.

Four appendices deal with issues that are loosely related to our main
argument, or with by-products of our analysis that may be interesting in their
own right. Appendix \ref{appA} motivates and presents a simple analytical
fit to the atomic data analyzed in Sec.~\ref{atoms}. Appendix \ref{appB} 
contains a comparison of the Lieb-Oxford bound with the Levy-Perdew bound,
appendix \ref{appC} classifies common (and some less common) parametrizations 
of the electron-liquid correlation energy with respect to the Lieb-Oxford 
bound, and appendix \ref{appD} discusses systems that violate the Lieb-Oxford 
bound.

\section{\label{atoms}Lieb-Oxford bound in atoms}

\begin{table}
\caption{\label{table1}
Exchange-correlation energy, LDA exchange energy, and their ratio
$\lambda$ for light atoms.
First three rows (He, Be and Ne): Near-exact numerical $xc$ and $x$
energies from QMC data, evaluated on exact densities.\cite{umrigar,burke}
Rows 4 to 20: Precise numerical $xc$ energies obtained\cite{goshparr} from CI
calculations and inversion of the KS equation by the ZMP procedure,\cite{zmp} 
and exchange-only LDA energies obtained self-consistently on LDA-PW92 densities.
For He, Be, and Ne these data are very similar to those obtained from QMC,
illustrating that the resulting values of $\lambda$ are robust, and not 
contaminated with inaccuracies due to approximate densities.\cite{footnote2} 
All values are in atomic (Hartree) units.}
\begin{ruledtabular}
\begin{tabular}{r|ccc}
     & $-E_{xc}$ & $-E_x^{LDA}$ & $\lambda$ \\
\hline
He & 1.067  & 0.8830   & 1.208  \\
Be & 2.770  & 2.321    & 1.193  \\
Ne & 12.48  & 11.02    & 1.132  \\
\hline
He & 1.068  &  0.8617  &  1.239  \\
Li & 1.827  &  1.514   &  1.207  \\
Be & 2.772  &  2.290   &  1.210  \\
B  & 3.870  &  3.247   &  1.192  \\
C  & 5.210  &  4.430   &  1.176  \\
N  & 6.780  &  5.857   &  1.158  \\
O  & 8.430  &  7.300   &  1.155  \\
F  & 10.320 &  8.999   &  1.147  \\
Ne & 12.490 &  10.967  &  1.139 \\
Na & 14.440 &  12.729  &  1.134 \\
Mg & 16.430 &  14.563  &  1.128 \\
Al & 18.530 &  16.486  &  1.124 \\
Si & 20.790 &  18.544  &  1.121 \\
P  & 23.150 &  20.743  &  1.116 \\
S  & 25.620 &  22.950  &  1.116 \\
Cl & 28.190 &  25.305  &  1.114 \\
Ar & 31.270 &  27.812  &  1.124
\end{tabular}
\end{ruledtabular}
\end{table}

In order to calculate $\lambda[n]$ we need the exact LDA exchange energy,
$E_{x}^{LDA}$, and the exact exchange-correlation energy $E_{xc}$, both on
the exact density. For a few closed-shell atoms, near-exact $E_{xc}$ values
and densities have been obtained by Umrigar and collaborators from QMC
calculations.\cite{umrigar} Of course, near-exact is not the same as
mathematically exact, but the margin of error of these QMC data is much
smaller than the effects we are after in this work.

The first three rows of Table \ref{table1} compare the near-exact $xc$ 
energies for He, Be and Ne (from Ref.~\onlinecite{umrigar}, as quoted in
Ref.~\onlinecite{burke}), the exact LDA exchange energies (obtained by
evaluating the LDA functional for exchange on exact QMC densities\cite{umrigar})
and the resulting ratio $\lambda[n]=E_{xc}/E_x^{LDA}$. Two trends
immediately leap to the eye: (i) the values of $\lambda$ are much
smaller than the theoretical upper limit $\lambda_{LO}=2.275$, and
(ii) $\lambda(Z)$ decreases as a function of atomic number $Z$.

To explore these emerging trends for a larger data set, the comparison is
extended in the other rows of Table \ref{table1} to other atoms.
For these atoms apparently no QMC results for $E_{xc}$ and the densities
are available. The exchange-correlation energies reported in rows 
4-20 were extracted from Ref.~\onlinecite{goshparr}, where they were obtained
by numerical inversion of the Kohn-Sham (KS) equation on CI densities,
following the Zhao-Morrison-Parr (ZMP) procedure.\cite{zmp} 
The values for $E_x^{LDA}$ in rows 4-20 were calculated
from the exact LDA exchange functional and evaluated at self-consistent
LDA(PW92) densities.\cite{footnote2} For He, Be and Ne, the resulting values
of $\lambda$ can be compared to those obtained from QMC. Cleary, both sets
of data differ slightly, but this difference is a small fraction of the 
difference between the observed $\lambda$ values and the theoretical upper 
bound $\lambda_{LO}$. Additionally, we have employed approximate xc energies
obtained from the B88-LYP GGA functional,\cite{b88,lyp} which is highly precise 
for atoms (and, unlike similarly precise nonempirical functionals, such as
PBE GGA and TPSS meta-GGA, does not make use of the Lieb-Oxford bound 
in its construction).

\begin{figure}[t!]
\includegraphics[height=80mm,width=90mm,angle=0]{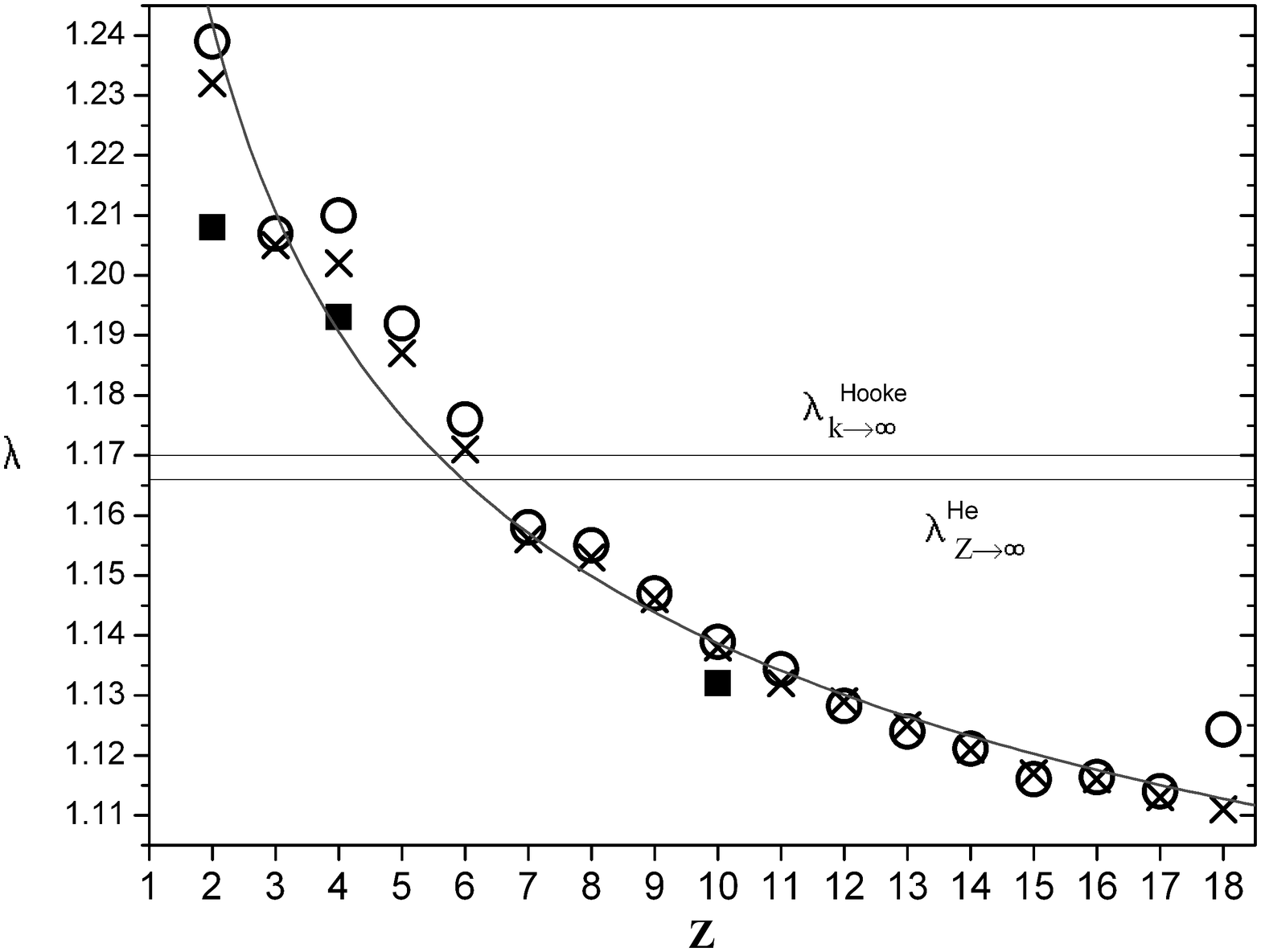}
\caption {\label{figure1} Value of the energy ratio $\lambda$, defined in
Eq. (\ref{lambdan}) as a function of atomic number $Z$ for light atoms.
Full squares: near-exact QMC data. Open circles: CI-ZMP data. Crosses:
supplementary approximate data obtained from the B88-LYP functional.
The upper limit $\lambda_{LO}$ is not shown in the figure,
because on this scale it is too far from the actual values of $\lambda$.
The two horizontal lines are estimates of $\lambda$ for Hooke's atom and for
the $Z\to\infty$ limit of Helium-like ions, discussed in Sec.~\ref{hooke}
and Sec.~\ref{ions}. 
The continuous line represents the analytical expression (\ref{atomfit}),
which is further discussed in Appendix~\ref{appA}.}
\end{figure}

Figure \ref{figure1} illustrates the simple and systematic trend of $\lambda$ 
as a function of $Z$, showing that the Lieb-Oxford bound with the originally 
proposed value $\lambda_{LO}=2.275$ is tightest for small $Z$, but actually
rather generous for all atoms, typical values of $\lambda$ being about 
$50\%$ smaller. Note also that the QMC data, which are expected to be more 
precise than the CI-ZMP data, systematically predict still smaller values 
of $\lambda$. For the Ar atom ($Z=18$), and perhaps also the Be atom ($Z=4$),
we suspect that the ZMP procedure has resulted in a less precise 
correlation energy than for the other atoms, because the correlation energy 
predicted for these atoms by the B88-LYP functional is much closer to the 
extrapolation of the QMC and other CI-ZMP data than their CI-ZMP values.

A simple fit to the B88-LYP data is
\be
\lambda(Z)= 0.993 + \frac{0.313}{Z^{1/3}},
\label{atomfit}
\ee
which is  represented by the continuous line in Fig.~\ref{figure1}.
A physical motivation for the form of this fit is given in Appendix~\ref{appA}.

From Table \ref{table1} and Fig. \ref{figure1} we conclude that 
for atoms $\lambda$ follows a simple and systematic trend as a function of $Z$,
and always remains far from the upper limit $\lambda_{LO}=2.275$, approaching
approximately half of this value as $Z\to 1$ and extrapolating to 1 as 
$Z\to \infty$. These conclusions are robust with respect to the various 
different ways of obtaining the densities and energies.\cite{footnote2}
Additional comparisons with the Levy-Perdew bound [second inequality of 
Eq.~(\ref{ineq})] are made in Appendix \ref{appB}.

\section{\label{hooke}Lieb-Oxford bound in Hooke's atom}

\begin{table}[t!]
\caption{\label{table2} 
Exact exchange-correlation and LDA exchange energies of Hooke's
atom, evaluated on the exact densities, for different values of the spring
constant $k$. Data for $k=0.25$ and $k=3.6\times10^{-6}$ (row one and two)
are from 
Ref.~\onlinecite{FiliUmTaut}. Alternative data for $k=0.25$ (row three) are 
from Ref.~\onlinecite{kais}, and the $k\to\infty$ limit has been extracted in
the way described in the main text, from data in Ref.~\onlinecite{laufer}.}
\begin{ruledtabular} 
\begin{tabular}{cccc}
 k   & $E_{xc}$ & $E_x^{LDA}$  &  $\lambda$ \\
\noalign{\smallskip}
\hline
$3.6\times10^{-6} $&  -0.0259 &  -0.0174    &  1.49 \\
0.25 &  -0.5536 &  -0.4410    &  1.255 \\
0.25 &  -0.555 &  -0.4410    &  1.26 \\
$k\rightarrow\infty$ & &          &  1.17
\end{tabular}
\end{ruledtabular}
\end{table}

Hooke's atom is a model system in which two electrons interact via Coulomb's 
law but are bound to a harmonic potential instead of a $1/r$ potential. It is 
frequently used in discussing approximate $xc$ functionals and other aspects 
of DFT, and many exact and numerical results for it are known.\cite{kestner,laufer,gill,kais,ivanov,taut,trickey}

The Hamiltonian of Hooke's atom is
\be
\hat{H}_{\rm Hooke}(k)=-\frac{\hbar^2}{2m}(\nabla_1^2+\nabla_2^2)
+ {1 \over 2} k (r_1^2 + r_2^2) + \frac{e^2}{|{\bf r}_1-{\bf r}_2|}.
\label{hookeham}
\ee
It describes two interacting electrons confined in space by a harmonic 
potential of strength $k$. Since the two electrons in this atom interact 
by the same Coulomb interaction $1/r_{12}$ as in real atoms, this system 
has the same $xc$ functional. The replacement of the nuclear potential by a
harmonic confinement, on the other hand, greatly simplifies the solution
of the eigenvalue problem posed by Eq.~(\ref{hookeham}) This fact has
motivated much work on this simple, yet nontrivial, model.\cite{kestner,laufer,gill,kais,ivanov,taut,trickey}

For the present purposes we are interested in the ratio $\lambda[n]$, which
we rewrite as
\be
\lambda[n]=\frac{E_x[n]}{E_x^{LDA}[n]}+\frac{E_c[n]}{E_x^{LDA}[n]}.
\label{hookelimit1}
\ee
Laufer and Krieger\cite{laufer} have shown that for $k\to \infty$ the LDA
exchange energy evaluated on the exact density recovers $85.5\%$
of the exact exchange energy of Hooke's atom. Hence, for large 
$k$ (high curvature),
\be
\frac{E_x^{LDA}}{E_x} = 0.855.
\label{laufer1}
\ee
As $k\to \infty$, the correlation energy of Hooke's atom rapidly drops to zero,
relative to the exchange energy. Hence, we can neglect the second term in 
Eq.~(\ref{hookelimit1}) and estimate $\lambda^{Hooke}$ from the inverse 
of Eq.~(\ref{laufer1}), which yields
\be
\lambda^{Hooke}_{k\to \infty} = 1.17,
\label{hookelimit2}
\ee
The value $\lambda_{k\to \infty}^{Hooke}$ is shown as a horizontal line in 
Fig.~\ref{figure1}.

Exact data at some finite values of $k$, including correlation, 
have been presented in Refs.~\onlinecite{kais} and \onlinecite{FiliUmTaut}. 
These data are collected in Table \ref{table2}.
The two independent calculations for $k=0.25$ are in excellent agreement, and
the tendency as a function of $k$ is consistent with our estimate of the 
$k \to \infty$ limit. Clearly, $\lambda$ for  Hooke's atom is very close
to its value for real atoms, and far below the limiting value $\lambda_{LO}$.

\section{\label{ions}Lieb-Oxford bound in ions from the Helium
isoelectronic series}

Near-exact numerical data for some representatives of the Helium isoelectronic 
series have been obtained from Hylleraas wave functions by Umrigar and 
Gonze,\cite{umrgonze} and are displayed in Table \ref{table3}. Laufer and 
Krieger\cite{laufer} also consider the large $Z$ limit of the Helium 
isoelectronic series, for which they find $E_x^{LDA}/E_x = 0.8577$. From 
this we obtain, by the same reasoning used for Hooke's atom, 
\be
\lambda^{He}_{Z\to \infty} \approx 1.166.
\label{heseries}
\ee
The value $\lambda^{He}_{Z\to \infty}$ is shown as a horizontal line
in Fig.~\ref{figure1}, and also included in Table \ref{table3}.

The trend of the data in Table \ref{table3} as a function of $Z$ is indeed
consistent with $\lambda^{He}_{Z\to \infty} \approx 1.166$. Interestingly, 
the value of $\lambda$ for the He isoelectronic series is very similar to 
that of Hooke's atom, in particular in the limit of very strongly confining 
external
potentials ($Z\to\infty$ and $k\to\infty$, respectively). For all values of
$Z$, including negative and positive ions, the resulting values of $\lambda$ 
are much smaller than $\lambda_{LO}$.

\begin{table}[t]
\caption{\label{table3} 
Near-exact exchange-correlation energies and LDA exchange energies of ions,
evaluated on the exact densities, from Ref.~\onlinecite{umrgonze}, and
the resulting value of $\lambda$. The $Z\to\infty$ limit is that of 
Eq.~(\ref{heseries}). All energies are in Hartree units.}
\begin{ruledtabular}
\begin{tabular}{cccc}
Z &  $E_{xc}$ &  $E_x^{LDA}$  &  $\lambda$ \\ 
\hline
1 (H$^{-}$)    &  -0.422893  &  -0.337    &  1.25 \\
4 (Be$^{2+}$)  &  -2.320902  &  -1.957    &  1.186 \\
10(Ne$^{8+}$)  &  -6.073176  &  -5.173    &  1.174 \\
80(Hg$^{78+}$) & -49.824467  & -42.699    &  1.1669 \\
$Z\to\infty$   &            &         &  1.166
\end{tabular}
\end{ruledtabular}
\end{table}

\section{\label{molecules}Lieb-Oxford bound in molecules}

Exchange-correlation energies for the silicon dimer and a few small 
hydrocarbons have been obtained by Variational Monte Carlo (VMC) techniques by
Hsing {\em et al.}\cite{hsing} These data, together with the resulting values
of $\lambda$, are recorded in Table \ref{table4}.

\begin{table}[t]
\centering
\caption{\label{table4} VMC exchange-correlation energies and LDA exchange 
energies of a few small molecules,\cite{hsing} and the resulting values of
$\lambda$. Last row: DMC data for bulk Si, from Ref.~\onlinecite{cancio}.}
\begin{ruledtabular}
\begin{tabular}{cccc}
 molecule & $E_{xc}$ & $E_{x}^{LDA}$ & $\lambda$ \\ 
\hline
  C$_2$H$_2$   & -3.840 & -3.428 & 1.120 \\
  C$_2$H$_4$   & -4.606 & -4.110 & 1.121 \\
  C$_2$H$_6$   & -5.367 & -4.760 & 1.128 \\
  Si$_2$       & -2.028 & -1.762 & 1.151 \\
\hline
  Si(bulk)     & -33.23 & -27.66 & 1.201
\end{tabular}
\end{ruledtabular}
\end{table}

The resulting values of $\lambda$ are quite similar to those obtained for
atoms. Interestingly, $\lambda$ of the hydrocarbons is smaller than that 
of the C atom, whereas $\lambda$ of the silicon dimer is a bit larger than 
that of the Si atom. Also, $\lambda$ for the hydrocarbons slowly grows as 
a function of the number of H atoms. If there is any trend as a function of 
electron number, it is very weak, as is demonstrated by the bulk limit, for 
which included in the Table the DMC value for bulk Si, from 
Ref.~\onlinecite{cancio}. 

Unfortunately, this data set may be too small to draw any reliable inferences 
from such trends. What is beyond doubt, however, is that the molecular data 
predict $\lambda$ values that are roughly as far away from the limit 
$\lambda_{LO}$ as previously found for atoms and ions.

\section{\label{liquid}Lieb-Oxford bound in the electron liquid}

The calculations of the previous sections show that for localized atomic,
ionic and molecular densities
the Lieb-Oxford bound is rather generous. The strength of the
original Lieb-Oxford argument, however, rests in the fact that is holds
for arbitrary densities, and not just for certain subsets. In order to
extend the investigation to a completely different class of densities we thus
now turn to spatially uniform systems. {\em A priori} there is no reason why 
one would expect similar values of $\lambda[n]$ to the ones found for 
localized density distributions, although the value for bulk Si, mentioned at 
the end of the previous section, strongly suggests so. 

The homogeneous electron liquid is, of course, of paramount importance for
DFT, as the reference system on which the construction of the LDA and many
GGAs and meta-GGAs are based. It is also of interest in its own right as a 
model for the conduction band of simple metals and as a many-body system in 
which effects of the particle-particle interaction can be studied without 
the simultaneous presence of complications due to inhomogeneity in the 
single-body potential.\cite{quantliq}

The per-particle exchange energy of the homogeneous electron liquid is 
\be
e_{x}(r_s)=-\frac{3}{4 \pi} \left(\frac{9 \pi}{4} \right)^{1/3} \frac{1}{r_s}
=:-\frac{D_0}{r_s},
\label{ex}
\ee
where $r_s$ is the usual electron-liquid parameter related 
to the charge density $n$ via $r_s=(4 \pi n/3)^{-1/3}$. Below we adopt the
value $D_0=0.4581653$, but note that the constants in the various 
parametrizations considered below are not normally known to this
number of significant digits.
For the electron liquid, and only for the electron liquid, the LDA is
by construction exact, and Eq. (\ref{lambdan}) becomes
\be
\lambda(r_s)=\frac{e_x(r_s)+e_c(r_s)}{e_x(r_s)} 
=1+\frac{e_c(r_s)}{e_x(r_s)}.
\label{ratio}
\ee
The per-particle correlation energy of the electron liquid is
not known in closed form, but the PW92 parametrization,\cite{pw92,primer}
\bea
e_c(r_s)=
-2 c_0(1+\alpha_1 r_s) \times 
\nonumber \\
\ln \left[ 1+\frac{1}{2c_0(\beta_1 r_s^{1/2} 
+\beta_2 r_s + \beta_3 r_s^{3/2} + \beta_4 r_s^2)}\right]
\label{ec}
\eea
is the best available fit to the Green's function Monte Carlo data of 
Ref.~\onlinecite{ceperley}.
In Eq. (\ref{ec}) $c_0=0.031091$, $\beta_1=7.5956$, 
$\beta_2=3.5875$, $\beta_3=1.6382$, $\beta_4=0.49294$ and $\alpha_1=0.21370$ 
are determined such as to reproduce the exactly known properties and the QMC 
data for $e_c(r_s)$.\cite{pw92,primer}

The unique combination of facts that the LDA becomes exact for the electron 
liquid and that $e_c$ is known to very high precision, allows us
to study $\lambda$ as a continuous function of the density parameter $r_s$,
instead of at isolated densities, as in the previous sections.

The high-density limit is, rigorously,
$\lambda(r_s\to0)=1$, because $\lim_{r_s\to 0} e_c(r_s)/e_x(r_s)=0$. 
To determine the low-density limit, recall the leading term of the 
large-$r_s$ expansion of the correlation energy,\cite{pw92}
\be
e_c(r_s\to \infty)= -\frac{d_0}{r_s}+\ldots,
\ee
where, according to best estimates \cite{pw92}, $d_0=0.43776$. This gives
the electron-liquid limit for $\lambda$,
\be
\lambda(r_s\to\infty) = 1+\frac{-d_0/r_s}{-D_0/r_s}=1.9555.
\ee
This limit is only of formal relevance, as at $r_s\approx 65\pm10$ the electron
liquid becomes unstable with respect to the Wigner crystal,\cite{ballone}
and the homogeneous phase ceases to be the ground state. Thus, the
largest physically possible value of the uniform electron liquid is 
$\lambda(r_s\approx 65)=1.65$. (This value changes only very little, if 
the older and presumably less accurate estimate\cite{ceperley} 
$r_s=100\pm 20$ for the liquid-to-crystal transition is used instead.)

The quantity $d_0$ can also be evaluated by employing in Eq.~(\ref{ratio}) 
available parametrizations of the electron-liquid correlation energy. 
From the PW92 parametrization, as specified above, one finds,
for example, 
\be
\lambda^{PW92}(r_s\to\infty) = 1+\frac{\alpha_1}{\beta_4}\frac{4 \pi}{3}
\left(\frac{4}{9 \pi}\right)^{1/3} = 1.9462.
\ee
A classification of common (and some less common) parametrizations of
the electron-liquid correlation energy with respect to the Lieb-Oxford
bound is presented in Appendix~\ref{appC}.

\begin{figure}
\includegraphics[height=80mm,width=90mm,angle=0]{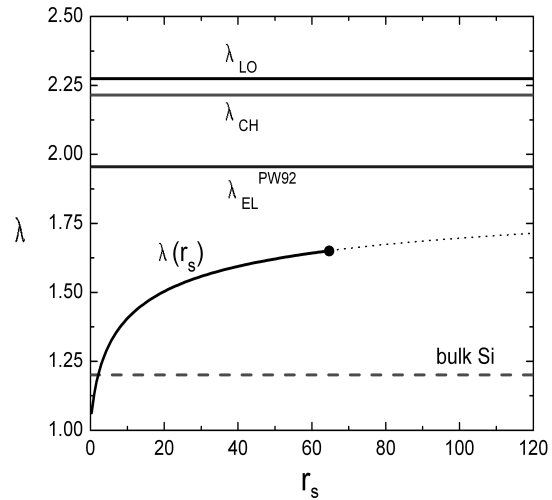}
\caption {\label{figure2} Value of the ratio $\lambda$, defined in Eq.
(\ref{lambdan}), as a function of the density parameter $r_s$ of the
homogeneous electron liquid, using the PW92 parametrization (\ref{ec})
for the correlation energy. The four horizontal lines correspond, from top 
to bottom, to the Lieb-Oxford\cite{lopaper} value $\lambda_{LO}$, the 
Chan-Handy\cite{handylo} value $\lambda_{CH}$, the low-density limit 
$\lambda_{r_s\to\infty}^{PW92}$, and the value for bulk $Si$, from 
Table~\ref{table4}. Note that actual metals are located at the very left 
margin of the plot, which we have extended to unphysically large $r_s$ 
values only to display the large remaining distance to $\lambda_{LO}$. 
The filled circle indicates the $r_s$ value where the transition to the 
Wigner crystal is expected to occur,\cite{ballone} and the uniform electron 
liquid ceases to be the ground state.}
\end{figure}

We note that the best estimate of $d_0$ still yields a value of 
$\lambda(r_s\to\infty)$ that is substantially smaller 
than the value $\lambda_{LO}$. Since the value $\lambda(r_s\to\infty)$ is
itself an upper limit of $\lambda(r_s)$ at all densities of the electron
liquid, this implies that in the entire range from $r_s=0$ to $1/r_s=0$
the Lieb-Oxford bound can be substantially tightened.

\begin{figure*}[t!]
\includegraphics[height=7cm,angle=0]{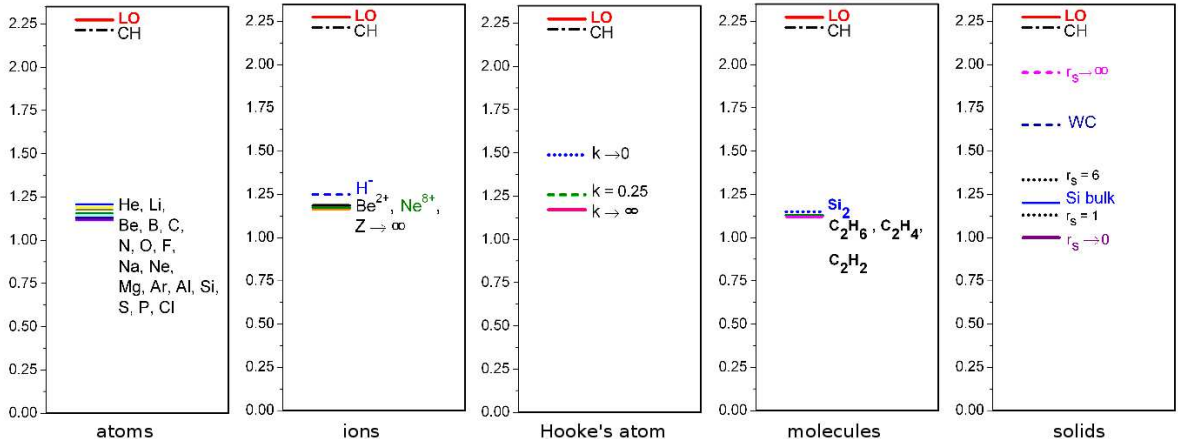}
\caption{\label{figure3} Graphical summary of all investigated systems.
Each panel shows, on the same scale, the Lieb-Oxford bound and its
refinement by Chan and Handy, together with actual values of $\lambda$
obtained from QMC or CI calculations, or exactly. The first panel shows that
the entire periodic table fits in a small stretch of $\lambda$ values bounded
from above by that of the $He$ atom. Ions, small molecules, Hooke's atom, and
solids -- represented by bulk Si and the electron liquid -- occupy almost the
same region of $\lambda$ space. Values above $\lambda=1.35$ are found only
for unphysical limits of model Hamiltonians ($r_s\to\infty$ of the electron
liquid, or $k\to0$ of Hooke's atom).}
\end{figure*}

This is illustrated in figure \ref{figure2}, which shows a plot of 
$\lambda(r_s)$ resulting from the PW92 parametrization over a wide density 
interval. Clearly, $\lambda(r_s)$ interpolates smoothly between the known 
limits, and remains far from its theoretical upper limit $\lambda_{LO}$ for 
any density. Figure~\ref{figure2} and the value $\lambda({r_s\to\infty}) = 
1.9555$ show that 
for any possible density of the electron liquid the Lieb-Oxford bound falls
way above the actual value of $\lambda$. For metallic densities, 
$r_s\approx(2\ldots 6)$, $\lambda$ is even smaller, not passing $1.333$. 

The $\lambda$ value of bulk Si, $\lambda=1.201$, included in Table 
\ref{table4}, falls near the center of the $\lambda$ interval obtained
for the electron liquid in the metallic density range, although bulk Si is 
not a metal. This again indicates that the shape of the density distribution 
is fairly unimportant for the value of $\lambda$, which never seems to come
even near the maximum $\lambda_{LO}=2.275$.

\section{\label{sum} Conclusions}

The two quantities, $\lambda[n]$ and $\lambda_{LO}$, considered in this paper 
have different meanings. $\lambda[n]$, as defined in Eq.~(\ref{lambdan}), 
measures the (inverse) weight of LDA exchange in the full exchange-correlation 
energy of an actual physical system, or a model of a physical system. 
We found here 
that across very different types of systems $\lambda[n]$ does vary, but not
very strongly. $\lambda_{LO}$, as defined in Eq.~(\ref{loalternat}), is a 
mathematical upper limit to $\lambda[n]$, which is the same for all 
nonrelativistic three-dimensional systems with Coulomb interactions.

The analysis of these two quantities, performed in the preceding sections,
can be summarized as follows: The Lieb-Oxford bound provides a lower bound 
on the $xc$ energy of any (nonrelativistic three-dimensional Coulomb) system, 
but nature does not necessarily make 
use of the entire range of permitted values, up to $\lambda_{LO}=2.275$.
Neutral atoms have $\lambda$ near $1$ and approach $1$ more closely as the 
atomic number $Z$ is increased. In this limit LDA exchange thus captures 
a larger part of the full $xc$ energy. Positive and negative ions,
Hooke's atom, small molecules and bulk Si all have $\lambda$ values that 
are close to those of isolated atoms, and far from $\lambda_{LO}$.
The electron liquid at metallic densities has $\lambda$ close to 
$1.3$, approaching $1$ and $1.65$ in the extreme high and low-density 
limits, respectively. 

Figure~\ref{figure3} contains a graphical summary of our analysis. Fine
details are not visible on the scale of the figure, but the overall impression
is very clear, and strongly suggests that the LO bound is too generous.
A tendency that systems with a more dilute and spread-out density
distribution produce larger $\lambda$ values is clearly visible for Hooke's
atom and the electron liquid. For atoms and ions, we find that lighter 
systems produce larger $\lambda$ values. For all investigated systems,
$\lambda$ is much smaller than $\lambda_{LO}=2.275$. Situations leading
to the, relatively, largest $\lambda$ values can be classified as follows:

(i) The largest values of $\lambda$ we found in this investigation arise 
in unphysical low-density limits of model Hamiltonians (the $k\to 0 $ limit of 
Hooke's atom, or the $r_s\to\infty$ limit of the electron liquid). The largest 
value we have found including such limits is $\lambda_{max}= 1.9555$, for 
the unattainable low-density limit of the electron liquid.

(ii) If we exclude physically unrealizable limits of a parameter approaching 
infinity or zero, the largest value we have found for physically possible
parameters is $\lambda_{max}=1.65$, for the very low-density uniform electron 
liquid, right at its transition to the Wigner crystal.

(iii) The largest value we have found for any actual physical system (atoms,
ions, molecules, and solids, but excluding Hooke's atom and the electron 
liquid, which are idealized models) is $\lambda_{max}=1.25$, for the H$^-$ ion.

This sequence of observations suggests two conjectures.

{\em Strong conjecture: Tightening the bound for all densities.} 
The Lieb-Oxford bound can be substantially tightened, in the mathematical 
sense. Independent support for this conjecture
comes from recent numerical work of Chan and Handy,\cite{handylo} who
indeed found that $\lambda_{LO}$ can be replaced by the slightly smaller
value $\lambda_{CH}=2.2149$. The present investigation suggests, however,
that a much larger reduction may be achievable.

It is, of course, conceivable that for certain special densities $\lambda[n]$ 
comes arbitrarily close to $\lambda_{LO}$ (or $\lambda_{CH}$), 
but if this is not the case for systems as different as uniform electron 
liquids and isolated atoms and ions, and moreover, if the distance from
the actual $\lambda$ in these systems to $\lambda_{LO}$ is a 
large fraction of $\lambda_{LO}$ itself, it becomes a very real possibility 
that $\lambda_{LO}$ is not the tightest possible limit. Still, if it should
turn out to be possible to construct (perhaps pathological) density 
distributions that require keeping the upper limit at $\lambda=\lambda_{LO}$
(or $\lambda_{CH}$), this would still be compatible with the following weaker 
conjecture.

{\em Weak conjecture: Tightening the bound for physical densities.} For 
physical densities (arising from realistic Hamiltonians, excluding unphysical 
limits) the Lieb-Oxford bound 
can be substantially tightened. In fact, our collection of data on different
systems suggests a limit of $\lambda_{conjec}\approx 1.35$, instead of
$\lambda_{LO}=2.275$. Interestingly, and somewhat suggestively, this 
empirically found value of $\lambda$ makes the prefactor $C$ in $E_{xc}[n] 
\ge -C \int d^3r\,n^{4/3}$ equal to unity.

Many existing density functionals, such as the PBE GGA \cite{pbe} and the 
TPSS meta-GGA \cite{tpss} employ the Lieb-Oxford bound, with 
$\lambda_{LO}=2.275$, as a constraint in their construction. It should be 
most interesting to explore if a change of $\lambda$ to a somewhat lower 
value, either universally or for classes of systems, 
has an impact on the performance of these functionals in applications in
electronic-structure calculations.

{\bf Acknowledgments}
This work was sup\-por\-ted by FAPESP and CNPq. Useful conversations with
Irene D'Amico and H.~J.~P. Freire are gratefully acknowledged.

\begin{appendix}

\section{\label{appA}Fit to the atomic data}

The systematic trend displayed by the atomic data in Fig.~\ref{figure1}
suggests that their behaviour follows a simple law as a function of $Z$. 
A reasonable form for this law can be guessed as follows: The correct value 
of $\lambda(Z\to \infty)$ is probably exactly one, because asymptotically 
exchange dominates correlation, and LDA exchange becomes exact.\cite{pcsb} 
For finite $Z$, we recall
that in generalized Thomas-Fermi theory\cite{spruch} all energy contributions 
are expanded powers of $Z^{1/3}$, so that the energy ratio $\lambda$ should 
also involve such powers. Since $\lambda(Z)$ decreases with increasing $Z$, 
the simplest expression consistent with these expectations is
\be
\lambda(Z)= A_1 + \frac{A_2}{Z^{1/3}},
\ee
where $A_2$ is a prefactor that cannot be determined by such generic
considerations, and we allow $A_1$ to deviate slightly from unity, because 
the available data points are not exact.

A simple fit of this expression to the data in Fig.~\ref{figure1} predicts 
$A_1=1.03$ and $A_2=0.231$ if only the three QMC data points are considered, 
$A_1=0.986$ and $A_2=0.331$ for the CI-ZMP data (except Ar, which, as explained 
above, may not be properly represented by the CI-ZMP data), and 
$A_1=0.993$ and $A_2=0.313$ for the approximate B88-LYP data, up to $Ar$. 
Reasuringly, the fitted values of $A_1$ are close to the theoretical
expectation $A_1=1$. The third of these fits is shown as continuous curve 
in Fig.~\ref{figure1}.

This simple fit already accounts well for the data in Fig.~\ref{figure1}, 
suggesting that the proposed $Z^{-1/3}$ behaviour is quite realistic. Better 
fits could, of course, be obtained by allowing more terms in the fitting 
function. However, our aim here is not to obtain the best possible fit to 
the data points but to illustrate that for atoms the prefactor $\lambda$ has 
a simple and systematic trend as a function of $Z$. 

\section{\label{appB}Levy-Perdew bound in atoms}

\begin{table}[t]
\caption{\label{table5} Comparison of the Levy-Perdew bound and the Lieb-Oxford
bound. Values of $B$ are taken from Ref.~\onlinecite{lp93}, where they are
obtained from a modified PW91 functional, evaluated at Hartree-Fock densities.
The values for $\lambda$ of He and Ne are QMC data from Table~\ref{table1}. 
For Ar and Kr, Ref.~\onlinecite{lp93} also provides values for $B$, but for 
Kr there do not seem to exist QMC or CI-ZMP data, and for Ar the CI-ZMP value 
is so far off the curve describing the other atoms (see Fig.~\ref{figure1}) 
that we do not trust it. Hence, we evaluated $\lambda$ for Ar and Kr 
selfconsistently from the B88-LYP
functional, which is highly precise for atoms. To illustrate the error that
arises from using an approximate functional, we have also included, in
row 3, the value of $\lambda$ predicted by the B88-LYP functional for Ne.}
\begin{ruledtabular}
\begin{tabular}{l|cc}
     & $\lambda$ & $B/E_x^{LDA}$ \\
\hline
He & 1.208 & 1.900 \\
Ne & 1.132 & 1.920 \\
\hline
Ne & 1.138 & \\
Ar & 1.111 & 1.926 \\
Kr & 1.079 & 1.931
\end{tabular}
\end{ruledtabular}
\end{table}

For the He, Ne, Ar and Kr atoms we can compare the Lieb-Oxford to the
Levy-Perdew bound [second inequality of Eq.~(\ref{ineq})]. 
Our analysis is based on the data in Table 1 of Ref.~\onlinecite{lp93},
which reports values of the quantity $B[n]=
\lim_{\gamma\to 0} \gamma^{-1}E_{xc}[n_\gamma]$ for one of the best available
GGAs (which at the time of writing of Ref.~\onlinecite{lp93} was a slightly
modified PW91\cite{pw91}), evaluated on tabulated
Hartree-Fock densities. To the extent that the PW91 functional and the
tabulated Hartree-Fock densities can be trusted as approximations to the exact
ones, the quantity $B^{PW91}[n^{HF}]$ should, according to
Eq.~(\ref{ineq}), provide a tighter bound on $E_{xc}$ than the Lieb-Oxford 
inequality, as indeed it was found to do.\cite{lp93} 

The corresponding values of the ratio
$B^{PW91}[n^{HF}]/E_{x}^{LDA}[n^{HF}]$ should thus be smaller than
$\lambda_{LO}$ but still larger than the actual ratio $\lambda$, defined in
our Eq.~(\ref{lambdan}), thus leading to the chain of inequalities
\be
\frac{E_{xc}}{E_x^{LDA}} = \lambda \le \frac{B}{E_x^{LDA}} \le \lambda_{LO}.
\label{ineq2}
\ee

The data in Table~\ref{table5}
show that this expectation is bourne out, and that even the Levy-Perdew
bound is still considerably above the actual (near exact) value of $\lambda$.
Interestingly, the tendencies of $\lambda$ and of $B/E_x^{LDA}$ as functions
of $Z$ are opposite, the former decreasing and the latter increasing.

\section{\label{appC}Electron-gas correlation energy}

We recall from Sec.~\ref{liquid} that the electron liquid displays the
largest values of $\lambda$ in the low-density limit, $r_s\to \infty$, where
\be
\lambda= 
1+\frac{e_c}{e_x} \to 
1+\frac{d_0}{\frac{3}{4\pi}\left(\frac{9\pi}{4}\right)^{1/3}} =
1.9555.
\label{lambdael}
\ee

Many interpolations and parametrizations of $e_c(r_s)$ have been proposed
over the years. In Table~\ref{table6} we list the values of $d_0$ and
$\lambda(r_s\to \infty)$ predicted by Wigner's original interpolation
formula (W),\cite{wigner} the modification of Wigner's expression
by Brual and Rothstein (BR);\cite{BrualR} the parametrizations of
Gunnarsson and Lundqvist (GL)\cite{GunnarsonL} and von Barth and Hedin
(vBH),\cite{vBH} which are based on perturbation theory; those of
Vosko, Wilk and Nusair (VWN),\cite{VWN} Perdew and Zunger (PZ81),\cite{PZ81} 
and Perdew and Wang (PW92),\cite{pw92} based on Monte Carlo data; a simple
electrostatic estimate presented in Ref.~\onlinecite{primer}, and the 
recent proposal by Endo {\em et al.} (EHTY),\cite{ehty} which was specifically 
designed for the $r_s\to 0$ limit.

\begin{table}[t]
\caption{\label{table6} Coefficient $d_0$ and ratio $\lambda$ for 
common, and some less common, parametrizations of the electron-liquid
correlation energy.}
\begin{ruledtabular}
\begin{tabular}{ccc}
Funcional & $d_0$ & $\lambda(r_s\to \infty)$  \\ 
\hline
W       & 0.44000 & 1.9604 \\
BR      & 0.02890 & 1.0631 \\
GL      & 0.28472 & 1.6214 \\
vBH     & 0.56700 & 2.2375 \\
VWN     & 0.41433 & 1.9043 \\
PZ81    & 0.42681 & 1.9316 \\
PW92    & 0.43352 & 1.9462 \\
EHTY    & 1.1189  & $\infty$ \\
elstat.est. & 0.90000 & 2.9644 \\
\hline
near-exact & 0.43776 & 1.9555 
\end{tabular}
\end{ruledtabular}
\end{table}

The entries in Table~\ref{table6} fall in three classes. First,
the electrostatic estimate and the EHTY parametrization violate 
the Lieb-Oxford bound even in its most generous universal form, employing
$\lambda_{LO}$.
This is not a surprise, considering the crudeness of the electrostatic estimate
and the fact that the EHTY parametrization was designed to work well in the
$r_s\to 0$ limit, not the $r_s\to\infty$ limit.
Second, the von Barth-Hedin parametrization and Wigner's interpolation formula 
obey the Lieb-Oxford bound in its universal form, but violate the stricter
electron-liquid limit, which a proper LDA must also obey. Third, all other 
functionals are consistent also with this stricter limit. We do not recommend 
the use of any xc functional from the first or second class. 
The BR functional,
which was fitted to data on the He atom\cite{BrualR} is a special case, as it
predicts a value of $d_0$ that obeys all bounds, but is way off the best
available value. Hence, we do not recommend the use of this expression
for extended systems.

\section{\label{appD}Systems violating the Lieb-Oxford bound}

Elsewhere in this paper we have repeatedly refered to the Lieb-Oxford 
bound as {\em universal}. This use of the concept of universality is the same
commonly employed in DFT: a universal quantity (such as the Hohenberg-Kohn 
functional $F_{HK}$) or property (such as the Lieb-Oxford bound) is one that 
is the same for all systems that share common kinetic-energy and 
interaction-energy operators. In particular, such quantities or relations
are independent of the external potentials. Since the Lieb-Oxford bound is
universal in this sense, we could confront it, above, with data on a wide
variety of different systems.

The kinetic-energy operator changes, e.g,  in relativistic quantum mechanics.
Even in nonrelativistic quantum mechanics it changes if the dimensionality 
is reduced. We stress that the Lieb-Oxford bound was derived for 
nonrelativistic three-dimensional systems,\cite{lopaper} and it is not clear 
if similar results hold in two or one dimensions, or relativistically. If 
similar bounds can be shown to hold, we expect that the prefactor 
$\lambda_{LO}$ (or $C_{LO}$) will be different.

The interaction-energy operator changes, e.g., when DFT is applied to
model Hamiltonians.\cite{baldalett,confined,hemoprb,magyar} 
An interesting example
is the Hubbard model, where the interaction is local (acting only between 
electrons at the same site) and spin-selective (acting only between electrons 
of opposite spins). 

For fermions, the wave function of the Hubbard model is properly 
antisymmetrized, but this has no consequences for the energy, i.e., the
exchange energy is rigorously zero. The local-density approximation for
the Hubbard model, constructed in Ref.~\onlinecite{baldalett} and applied, 
e.g., in Refs.~\onlinecite{mottlett,superlattice,coldatoms}, respects this
property. Hence, the Lieb-Oxford bound in its Coulomb-interaction form
\be
E_{xc}[n] \ge \lambda_{LO} E^{LDA}_x[n]
\ee
cannot hold for the Hubbard model, because the right-hand side is rigorously
zero, whereas the left-hand side is known to be nonzero and negative. 
Similar conclusions hold for other model Hamiltonians whose interaction is
not of Coulomb form.

\end{appendix}

\end{document}